\documentclass[aps,prl,twocolumn,groupedaddress,showpacs]{revtex4}

\usepackage{graphicx,amsmath}
\usepackage{color}

\begin{document}

\title{Multispinon continua at zero and finite temperature in a near-ideal Heisenberg chain}
\author{
B. Lake$^{1,2}$, 
D. A. Tennant$^{1,2}$, J.-S. Caux$^3$, T. Barthel$^4$, U. Schollw\"ock$^4$, S.E. Nagler$^5$ and C.D. Frost$^6$}

\affiliation{
$^1$ Helmholtz-Zentrum Berlin, Hahn-Meitner Platz 1, D-14109 Berlin, Germany  \\
$^2$ Institut f{\"u}r Festk{\"o}rperphysik, Technishes Universit\"at Berlin, Hardenbergstr 36, 10623 Berlin, Germany \\
$^3$ Institute for Theoretical Physics, Universiteit van Amsterdam,
Science Park 904, Amsterdam, The Netherlands \\
$^4$ Department of Physics and Arnold Sommerfeld Center for Theoretical Physics,
Ludwig-Maximilians-Universit{\"a}t M{\"u}nchen, Theresienstr.\ 37, 80333 Munich, Germany\\
$^5$ Quantum Condensed Matter Division, Oak Ridge National Laboratory, Oak Ridge, TN, U.S.A. \\
$^6$ ISIS Facility, Rutherford Appleton Laboratory, Chilton OX11 0QX, United Kingdom
}

\begin{abstract}
The space- and time-dependent response of many-body quantum systems is the most informative aspect of their emergent behaviour. 
The dynamical structure factor, experimentally measurable using neutron scattering, can map this response in wavevector and energy with great detail, 
allowing theories to be quantitatively tested to high accuracy. Here, we present a comparison between neutron scattering measurements on the one-dimensional spin-1/2 Heisenberg antiferromagnet $\mbox{KCuF}_3$, and recent state-of-the-art theoretical methods based on integrability and density matrix renormalization group simulations. 
The unprecedented quantitative agreement shows that precise descriptions of strongly correlated states at all distance, time and temperature scales are now possible, and highlights the need to apply these novel techniques to other problems in low-dimensional magnetism. 
\end{abstract}
\maketitle

Understanding the emergent properties of many-body quantum states is a 
central challenge of condensed matter physics. Their response behaviour, encoded 
in dynamical structure factors, carries all the intricacies of strong correlations 
even in relatively simple systems. In this context, one-dimensional (1D) magnets have long been
a preeminent laboratory for developing new, more widely-applicable approaches and methods, as well
as for seeking correspondence between theory and experiments \cite{GiamarchiBOOK}.
The prototypical example is the 
spin-1/2 (S-1/2) Heisenberg antiferromagnet (HAF), described by the Hamiltonian 
\cite{1928_Heisenberg_ZP_49}
\begin{equation}
H = J \sum_j {\bf S}_j \cdot {\bf S}_{j+1}
\label{eq:HXXX}
\end{equation}
with nearest-neighbor exchange interaction $J$. Equation (\ref{eq:HXXX}) embodies the
combined challenges of non-linearity, embedded in the spin commutation relations, and strong ground-state 
fluctuations due to the small spin value. This model was first tackled in 1931 by Hans Bethe who 
obtained its eigenstates with the Bethe Ansatz \cite{1931_Bethe_ZP_71}. 
A full understanding of this model has since then remained one of the long-standing problems of condensed matter physics. 
The complexity of its eigenstates has meant that, to date, only a partial understanding of the dynamical response 
has been achieved which fails to provide sufficient accuracy or coverage to be able to establish the 
behaviour over the relevant time and distance scales. 

Here, we bring together crucial advances in theory, based on integrability and 
time-dependent density matrix renormalization group methods, to give a quantitative description of the 
dynamical response of the 1D S-1/2 HAF 
which we compare with the measured spectra of a model material. 
These new methods provide accurate correspondence over the entire observable parameter range, including finite temperatures, 
and allow for an unambiguous diagnosis of discrepancies in both experiments and other approximate theoretical approaches.
The combined insights from these techniques have wide applicability and provide 
a more general quantitative understanding of the quantum properties of strongly correlated 1D systems.

\begin{figure}[t]
\includegraphics[width=0.4\textwidth]{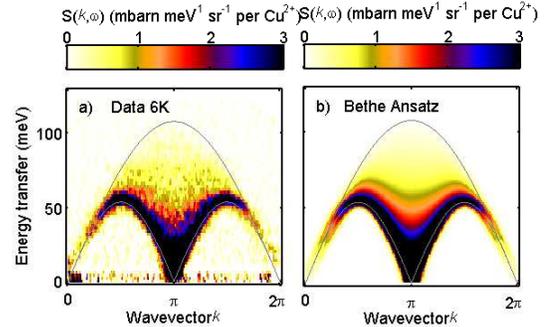}
\caption{\label{fig:fig_1} (color online) INS data compared to theory. a) The 
data show the multi-spinon continuum lying predominantly between the upper 
($\omega_u\left(k\right)$) and lower ($\omega_l\left(k\right)$) boundaries for 2-spinon processes (grey 
lines). b) The dynamical structure factor computed via the algebraic Bethe Ansatz.}
\end{figure}

{\it Experimental details and methods --} 
To compare the different theoretical approaches with high-accuracy data, we performed inelastic neutron scattering on the prototypical 1D S-$1/2$ HAF 
KCuF$_3$. In this compound, orbital order \cite{Lee2012} provides strong Heisenberg coupling ($J=33.5$meV) 
between the Cu$^{2+}$ (S-$1/2$) ions in the $c$-direction. KCuF$_3$ has a long history in the study of the 
1D S-$1/2$ HAF. The energy and wavevector dependence of the characteristic spinon continuum 
\cite{1991_Nagler_PRB_44,1993_Tennant_PRL_70,Tennant1995_2,Tennant1995_1}, as well as the presence of 
energy/temperature scaling indicating proximity to the Luttinger liquid quantum critical point were established for the first time in KCuF$_3$ \cite{Lake2005_1}. 
In addition, weak interchain coupling was shown to modify the low energy spectrum and, below the N\'{e}el 
temperature of $T_N=39$K, the existence of low energy spin-waves and a novel longitudinal mode were 
found to accompany symmetry breaking 
\cite{Lake1997_1,Essler1997,Lake2000_1,Lake2005_2,Lake2005_3}. For energies greater than $30$meV, the behavior 
of KCuF$_3$ is entirely 1D.

Inelastic neutron scattering (INS) is the most powerful method to analyze magnetic dynamics because the measured 
cross-section yields the dynamical structure factor (DSF) as a function of momentum $k$ and energy $\hbar\omega$,
\begin{equation}
S^{ab} (k, \omega) = \frac{1}{N}\sum_{j,j'}^N e^{-i k (j - j')} \int_{-\infty}^{\infty} dt e^{i\omega t} \langle S_{j}^a (t) S_{j'}^b(0) \rangle,
\label{eq:DSF}
\end{equation}
where $N$ is the number of sites, and $a,b = x,y,z$. 
INS datasets of KCuF$_3$ were collected at temperatures $T=6$K, 50K, 75K, 150K, 200K, 300K using the MAPS spectrometer at the ISIS 
Facility, Rutherford Appleton Laboratory, U.K. To make comparison with theory, we simulated the experiment based on the theoretical 
$S(k,\omega)$ (see supplemental material \cite{supplemental}). Figure 1a shows the experimental DSF for $T=6$K as a function of 
momentum along the chain and energy. It reveals the characteristic multi-spinon continuum of the 1D S-$1/2$ HAF and is in excellent 
agreement with the Bethe Ansatz solution for $T=0$K (fig.\ 1b) described later. Since the data is normalised to absolute units no 
overall scale factor was required when comparing theory and experiment.

\begin{figure}[t]
\includegraphics[width=0.30\textwidth]{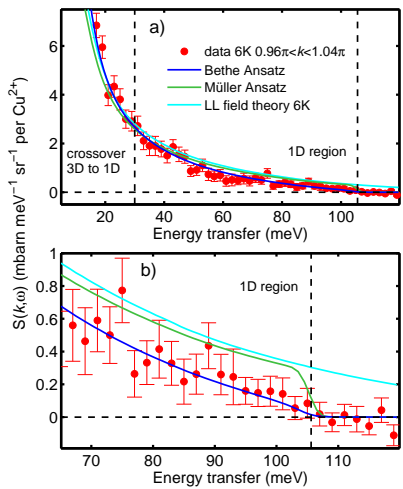}
\caption{\label{fig:fig_2} (color online) Comparison of the INS data at $k=\pi$ and $T=6$K, 
with the theoretical approaches. a) The data agrees approximately with the Luttinger liquid, 
M\"uller Ansatz and algrebraic Bethe Ansatz. b) Differences between 
the theories increase at higher energies and the Luttinger liquid and M\"uller Ansatz 
show strong discrepancies with the data near the 2-spinon upper threshold. 
}
\end{figure}

{\it Previous theoretical approaches --}
The ground state of the 1D S-$1/2$ HAF is quantum disordered with power-law correlations. The 
important low-lying excitations \cite{1962_desCloizeaux_PR_128} which define the DSF (\ref{eq:DSF}) 
are known as spinons \cite{1981_Faddeev_PLA_85} and can be pictured as N\'eel domain walls dressed by quantum 
fluctuations. They carry fractional spin (S-1/2) which restricts them to being created in (multiple) pairs, and 
they disperse according to $\frac{\pi}{2} J |\sin k|$. The simplest 
observable continuum, made from two spinons, fills the region $\omega_l (k) \leq \omega \leq \omega_u (k)$ 
between lower and upper boundaries
\begin{equation}
\omega_l (k) = \frac{\pi}{2} J |\sin k|, \,\,\,\, 
\omega_u (k) = \pi J |\sin \frac{k}{2}|, \,\,\,\,
k \in [0, 2\pi].
\label{eq:omegalu}
\end{equation}
The four-spinon continuum also has a lower threshold at $\omega_l (k)$ as do arbitrary $2n$-spinon states.

Because of the long-term absence of precise calculations for the DSF of the 1D S-1/2 HAF, finite-size exact diagonalization results, sum rules, and the spinon dispersions were combined into a phenomenological formula at $T=0$K - the so-called M{\"u}ller Ansatz (MA) \cite{1981_Mueller_PRB_24},
\begin{equation}
S_\text{MA} (k, \omega) = A_\text{MA} \frac{\Theta (\omega - \omega_l (k)) \Theta (\omega_u(k) - \omega)}{[\omega^2 - \omega_l^2 (k)]^{1/2}},
\label{eq:MA}
\end{equation}
where $A_\text{MA}=289.6/ \pi$. This formula, though historically important due to its simplicity, is inexact. 

Bosonization \cite{GiamarchiBOOK} can also be used to approximate the DSF at $k = 0,\pi$, where the spinon dispersion is linear and the system can be described as a Luttinger liquid (LL) \cite{1981_Haldane_JPC_14}. 
Finite temperatures are then straightforwardly treated, giving the DSF around 
$k = \pi + \delta k$ as \cite{1986_Schulz_PRB_34}
\begin{multline}
S_\text{LL} (\pi + \delta k, \omega, T) = \mspace{285mu} \\
\frac{e^{\frac{\hbar \omega}{kT}}}{e^{\frac{\hbar \omega}{kT}}-1} \frac{A_\text{LL}}{T} \operatorname{Im} \left[ \rho \left( \frac{\omega + v_F \delta k}{4 \pi T} \right) \rho \left( \frac{\omega - v_F \delta k}{4 \pi T} \right) \right],
\label{eq:Schulz}
\end{multline}
where $\rho (x) \equiv \Gamma (1/4 - i x)/\Gamma(3/4 - ix)$, $v_{F} = \frac{\pi}{2} J$ is the Fermi velocity and $A_\text{LL}$ is a constant. This approach is not applicable at generic momenta. 

Recent work making use of nonlinear LL theory \cite{2009_Imambekov_PRL_102,2009_Imambekov_SCIENCE_323} allows the threshold behaviour at all $k$ to be obtained for $T = 0$K \cite{2008_Pereira_PRL_100,2009_Pereira_PRB_79}. 
At $k = \pi$ and low energies it becomes a power law with logarithmic corrections, 
$S(k=\pi, \omega \rightarrow 0) \sim \frac{1}{\omega} \sqrt{\ln \frac{1}{\omega}}$, changing at $k \neq \pi$ to 
$S(k \neq \pi, \omega \rightarrow \omega_l (k)) \sim \frac{1}{\sqrt{\omega - \omega_l (k)}} \sqrt{\ln \frac{1}{\omega - \omega_l(k)}}$ for frequencies $\omega$ close to the lower threshold $\omega_l(k)$. There is no obvious extension of this result to finite temperatures.

\begin{figure}[t]
\includegraphics[width=0.275\textwidth]{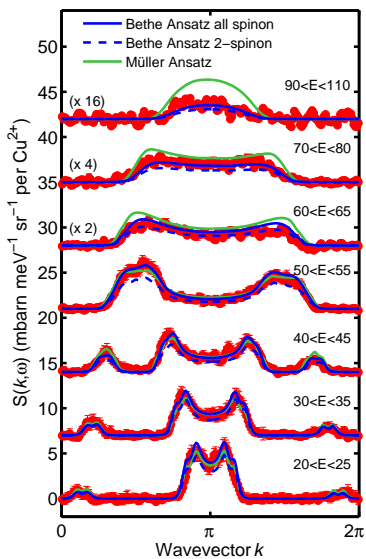}
\caption{\label{fig:fig_3} (color online) Comparison of the theories with the data at 6K as a function of 
wavevector at different energies. The M\"uller Ansatz strongly differs from the data above $55$meV. The relative 
importance of multispinon processes can be determined using the Bethe Ansatz (VOA). The 2-spinon process alone 
clearly underestimates the scattering, highlighting the importance of including higher-spinon terms.}
\end{figure}

{\it Comparison of these theories for $T = 0$K to the lowest temperature ($T = 6$K) KCuF$_3$ INS data --}
Figure 2 shows the data at $k=\pi$ and $T=6$K as a function of energy compared to the M\"uller Ansatz and 
Luttinger liquid theory at $T = 0$K. The measured intensity behaves approximately as the power law 
$S(\pi,\omega)\sim1/\omega^\eta$ ($\eta=1$) indicating proximity to the LL quantum critical point. Below 
$\approx 30$meV the correlations become increasingly modified from 1D to 3D due to interchain coupling, thus deviations
from 1D theories are expected. Above $30$meV however, KCuF$_3$ is completely dominated by 1D behavior. Both theories 
systematically overestimate the scattering at high energies showing clear quantitative differences from the data. 
The LL is a continuum field theory and is hence unable to capture the upper cutoff intrinsic to the finite lattice spacing; 
it thus becomes imprecise for higher energies. In the case of the MA the predicted stepwise upper cutoff at $\omega_u(k)$ is clearly incorrect. 
The MA is also inaccurate at other wavevectors. The constant energy cuts in Figure 3 show that it systematically 
overestimates the scattering for all energies above $55$meV and is inaccurate everywhere except near 
the lower continuum boundary. 

The cuts at different wavevectors ($k\neq \pi$) in Figure 4 reveal threshold singularities at the lower boundary $\omega_l(k)$ of the continuum. 
These are X-ray-edge-type singularities with power-law correlations extending to positive energies. 
Comparison to the nonlinear LL picture reveals that this theory is increasingly inaccurate as $k$ goes further from $\pi$. 
The linear LL picture can also be applied at finite temperatures although only in the region near $k=\pi$ \cite{1986_Schulz_PRB_34}. 
As shown in Fig.\ 5, where it is compared to higher temperature KCuF$_3$ data, it works approximately for temperatures up to $\approx$100K, 
but for larger temperatures more accurate descriptions are clearly necessary. 

{\it Bethe Ansatz approaches for $T=$0K --}
Two approaches based on the exact solvability of the 1D S-1/2 HAF Eq.\ \eqref{eq:HXXX} have recently provided much more reliable 
calculations of the ground state DSF. First, the Heisenberg model displays an emergent quantum group symmetry 
in the infinite system size limit $N \rightarrow \infty$. This is exploited in the vertex operator approach 
(VOA) \cite{JimboBOOK} to obtain eigenstates and matrix elements of spin operators. The 2-spinon contribution 
to Eq.\ (\ref{eq:DSF}) computed using the VOA \cite{1996_Bougourzi_PRB_54,1997_Karbach_PRB_55} yields $72.9\%$ 
of the integrated sum rule and $71.3\%$ of the first frequency moment \cite{1974_Hohenberg_PRB_10}. The 
remaining signal is carried by $4, 6, 8, ...$ spinon states. Matrix elements of spin operators between ground- and 4-spinon 
states \cite{1997_Abada_NPB_497} can be assembled into the 4-spinon contribution to the 
DSF \cite{2006_Caux_JSTAT_P12013}, remarkably yielding (along with 2-spinon parts) about $98\%$ of the DSF 
in the thermodynamic limit. 

%
\begin{figure}[t]
\includegraphics[width=0.35\textwidth]{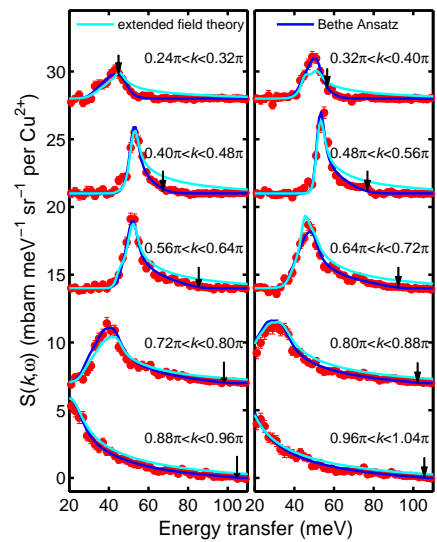}
\caption{\label{fig:fig_4} (color online) 6K data plotted as a function of energy for different wavevectors. 
Threshold singularities occur across the Brillouin zone at $\omega_l(k)$. These are compared to the algebraic Bethe Ansatz 
and the nonlinear Luttinger liquid theory. The black arrows indicate the 2-spinon upper threshold $\omega_u(k)$.}
\end{figure}
%
The second integrability-based approach
uses the algebraic Bethe Ansatz (ABA) \cite{KorepinBOOK}, exploiting exact
finite-size matrix elements of spin operators \cite{1999_Kitanine_NPB_554} which can be 
resummed \cite{2002_Biegel_EPL_59,2003_Biegel_JPA_36,2004_Sato_JPSJ_73} over relevant 
excitations \cite{2005_Caux_PRL_95,2005_Caux_JSTAT_P09003} 
({\it i.e.}\ arbitrary numbers of spinons) 
to obtain precise results for large systems (over $99\%$ saturation for 500 sites), for arbitrary 
field and anisotropy. The ABA and VOA methods give identical results (up to finite-size corrections and imperfect saturations) 
for the ground-state correlations. 

\begin{figure}[t]
\includegraphics[width=0.48\textwidth]{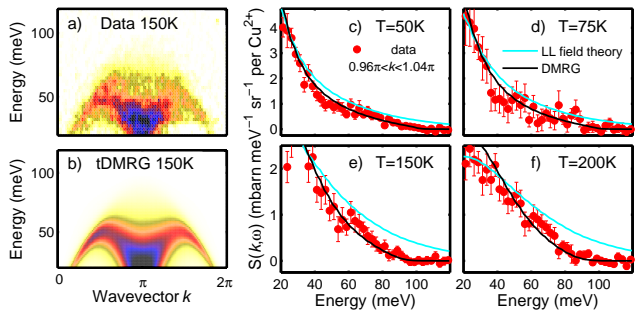}
\caption{\label{fig:fig_5} (color online) 
Finite temperature behavior. a) INS data at 150K is accurately described by, b), the tDMRG simulation. c-f) The temperature dependence at $k=\pi$. The Luttinger liquid theory agrees with the INS at low energies and temperatures. The tDMRG calculations however give precise agreement over the full energy and temperature range. Remaining differences at low energies are due to the interchain coupling and the background subtraction procedure. 
}
\end{figure}

{\it Comparison of the $T = 0$K Bethe Ansatz approaches to the lowest temperature ($T = 6$K) KCuF$_3$ INS data --} 
As shown in Figures 1-4 these theories provide excellent quantitative agreement over all energies and wavevectors. 
Unlike the M\"uller Ansatz and Luttinger liquid field theory, the integrability-based algebraic Bethe Ansatz at $k=\pi$ (Fig.\ 2) demonstrates the correct analytic 
behaviour at the 2-spinon upper boundary providing a quantitative description of the truncation of spinon states. 
The vertex operator approach is compared to the constant energy cuts (Fig.\ 3) and unlike the MA 
provides accurate agreement with the measurements throughout the Brillouin zone including at highest energies. The 
VOA can also be used to assess the relative importance of 2- and higher-spinon contributions to the 
scattering. 
Considering only 2-spinon processes (dashed line) shows marked differences from the measurements above $30$meV and away from 
$k=0,\pi$. 
Therefore, as suspected in Ref.~\cite{2010_Enderle_PRL_104}, and very recently shown in Ref.~\cite{2013_Mourigal_NATPHYS_9}, higher-order spinon processes must be included. 
Finally, unlike the nonlinear LL field theory, the Bethe Ansatz computations are able to capture the 
threshold singularities quantitatively throughout the Brillouin zone (Fig.\ 4). Furthermore they also agree with the 
cutoff from 2-spinon processes at the upper threshold, which is not a MA type step function but a 
square-root cusp.

{\it tDMRG for finite temperatures --}
The problem of the finite-temperature DSF remains for the moment inaccessible to these exact integrability-based methods. However, finite-temperature response functions of 1D systems, like $\langle S_{j}^a (t) S_{j'}^b(0) \rangle$ in Eq.~\eqref{eq:DSF}, can be evaluated in a quasi-exact manner up to some maximum reachable time $t_{\max}$ on the basis of the time-dependent density matrix renormalization group (tDMRG) \cite{Vidal2003-10,White2004,Daley2004}. A corresponding scheme, introduced in 
Ref.~\cite{Barthel2009-79b}, is based on a sequence of imaginary-time and real-time evolutions during which the occurring many-body operators are approximated in matrix product form. As described in Refs.~\cite{Barthel2009-79b} and \cite{White2008-77}, one can use linear prediction \cite{Yule1927-226,Makhoul1975-63} to extend the obtained data from the time-interval $[-t_{\max},t_{\max}]$ to infinite times before doing the Fourier transform in Eq.~\eqref{eq:DSF} that yields the DSF.
A difficulty in the DMRG simulations is the (typically linear) growth of entanglement with time \cite{Calabrese2005,Bravyi2006-97,Barthel2008-100}. In tDMRG calculations, this leads to a severe increase of the computation cost and strongly limits the maximum reachable times $t_{\max}$. It is only due to a novel much more efficient evaluation scheme for the thermal response functions \cite{Barthel2012_12,Barthel2013_01,Karrasch2013_03} that we are now able reach sufficiently large $t_{\max}$ such that the linear prediction becomes very accurate and, precise structure factors can be computed. 

{\it The tDMRG simulations compared to finite temperature KCuF$_3$ INS data --}
The results shown in Fig.\ 5, give the first application of this optimized tDMRG scheme \cite{Barthel2012_12,Barthel2013_01} to determine the full momentum- and energy-dependence of the DSF at $T>0$. The simulations were carried out with systems of 129 sites and a DMRG truncation weight \cite{Schollwock2005} of $10^{-10}$, guaranteeing negligible finite-size and truncation effects. The tDMRG results clearly provide an excellent description of the experimental cross-section without adjustable parameters except at lowest energies where the interchain coupling is significant. As mentioned before, the linear LL theory allows finite temperature comparison at $k=\pi$, however the assumption of a linear dispersion results in strong discrepancies at higher energies and temperatures. In contrast, tDMRG is able to accurately describe the system over the full energy and temperature range. It also provides an accurate description of the INS data throughout the Brillouin zone (not just at $k=\pi$ as for the LL theory). 

{\it Conclusion --} Detailed comparison to high-quality inelastic neutron scattering data shows the inadequacy of 
conventional approximations for the dynamic structure factor of the 1D S-$1/2$ HAF. Instead, excellent agreement 
is found with new theories based on exact solutions.
These comparisons directly show the importance of computing cross-sections beyond 2-spinon terms, and the correct fitting 
of the high-energy cutoffs.
Furthermore we have shown that the data at finite temperatures can be modeled by a novel DMRG method, giving excellent 
agreement over the full temperature, energy and wavevector range. This paper demonstrates that the combination of 
integrability and DMRG calculations provides a solution to the long-standing problem of the response of the 1D S-$1/2$ HAF over all experimental parameters. We anticipate that these powerful techniques will in the future be successfully applied to other problems in low-dimensional magnetism as they allow for unambiguous identification of deviations due to experimental phenomena \cite{Walters} and approximations in other theoretical approaches.

{\it Acknowledgements --}
J.-S. C. acknowledges NWO and the FOM foundation of the Netherlands. S.E.N. is supported by US DOE Basic Energy Sciences 
Division of Scientific User Facilities.


\end{document}